\documentclass[reprint,twocolumn,amsmath,amssymb,aps,prl,floatfix]{revtex4-2}

\usepackage{amsmath}
\usepackage{amssymb}

\usepackage{graphicx}               
\usepackage{wrapfig}

\usepackage{dcolumn}                

\usepackage{upgreek}                
\usepackage{bm}                     

\usepackage{xcolor}
\usepackage[colorlinks=true, linkcolor=blue, citecolor=blue, urlcolor=blue]{hyperref}
\usepackage{soul,xcolor}
\setstcolor{red}
\usepackage{accents}
\usepackage{wasysym}                

\usepackage{lineno}

\usepackage{mathrsfs}
\usepackage{comment}
\usepackage{mathtools}
\usepackage[version=4]{mhchem}
\usepackage{float}
\usepackage{enumitem}
\usepackage{tikz}
\usepackage{bibunits}
\usepackage{cancel}

\hypersetup{
    colorlinks=true,
    linkcolor=blue,
    filecolor=magenta,      
    urlcolor=blue,
    citecolor=blue,
}

\begin{document}

\title{Saddle Buckling of a Multicellular Alga Driven by Differential Growth}

\author{Zhao Wang}
\email{zw511@cam.ac.uk}
\affiliation{Department of Applied Mathematics and Theoretical Physics, University of Cambridge, Wilberforce Road, Cambridge CB3 0WA, United Kingdom}
\author{Sumit Kumar Birwa}
\email{skb61@cam.ac.uk}
\affiliation{Department of Applied Mathematics and Theoretical Physics, University of Cambridge, Wilberforce Road, Cambridge CB3 0WA, United Kingdom}
\author{Jane Y. Y. Chui}
\email{yyjc2@cam.ac.uk}
\affiliation{Department of Applied Mathematics and Theoretical Physics, University of Cambridge, Wilberforce Road, Cambridge CB3 0WA, United Kingdom}
\author{Kyriacos C. Leptos}
\email{kl344@cam.ac.uk}
\affiliation{Department of Applied Mathematics and Theoretical Physics, University of Cambridge, Wilberforce Road, Cambridge CB3 0WA, United Kingdom}
\author{Raymond E. Goldstein}
\email{R.E.Goldstein@damtp.cam.ac.uk}
\affiliation{Department of Applied Mathematics and Theoretical Physics, University of Cambridge, Wilberforce Road, Cambridge CB3 0WA, United Kingdom}

\date{\today}

\begin{abstract}
During development of many simple multicellular organisms 
there is a complex interplay between cell growth and 
generation of an extracellular matrix that confers structure. 
We study this problem quantitatively using \textit{Gonium pectorale}, a 16-cell alga that transforms from 
a flat to a 
buckled shape during growth without cell division or rearrangement.  
Cell-level measurements and 3D reconstructions show that the $12$ peripheral cells grow faster than the central
$4$, producing hoop stress that selects the saddle 
mode, whose amplitude displays a continuous bifurcation
with classical square root singularity. 
These results are consistent 
with a continuum model of differential growth within 
F\"oppl--von K\'arm\'an theory, which yields 
information on the mechanical properties of the ECM.
Implications of this transformation in evolutionary transitions to multicellularity are discussed.
 
\end{abstract}

\maketitle

In the long history of scientific studies of organism growth, it
was J.S. Huxley \cite{Huxley1932} 
who perhaps first clearly articulated the existence
of ``differential growth" as a key feature.  Inspired by
the earlier geometric analyses by D'Arcy Thompson
\cite{thompson1917growth}, 
across a wide range of 
animals Huxley provided empirical evidence not only
for organs growing at different rates within an organism 
(e.g., the claw of the fiddler crab relative to the rest of its
body),
but also for ``growth gradients" within a given organ (e.g., between
the proximal and distal regions of the claw).

Only much more recently have the mechanical implications
of differential growth on the three-dimensional shapes of 
continua been understood.  Within elasticity theory,
incompatible local growth rates create residual stress;
for ostensibly flat, sheet-like tissues, that 
stress can be relieved by out-of-plane bending 
~\cite{rodriguez1994stress,efrati2009elastic,goriely2005differential,dervaux2008morphogenesis,dervaux2009morphogenesis}, as in patterned gels~\cite{klein2007shaping,kim2012designing,mora2006buckling} and in growing leaves, flowers, and the gut~\cite{liang2009shape,liang2011growth,savin2011growth,nath2003genetic,coen2004genetics,zhang2026roses}.  The response of the tissue depends on
the spatial distribution of the growth and on the nature of
the mechanical properties of the tissue.

These issues take on particular significance when 
considering simple organisms
that serve as models for evolutionary transitions to 
multicellularity.  Both for Volvocine green algae 
(spanning from unicellular \textit{Chlamydomonas} to 
\textit{Volvox}) and choanoflagellates (the closest living
relatives of animals), cells are typically held together
by an extracellular matrix (ECM).  As this
provides the physical scaffold that dictates overall 
morphology, and directly enables specialized behaviours like swimming, feeding, and phototaxis, proper 
orchestration of cell growth and ECM generation is 
crucial.  In higher organisms, tissue growth is 
often accompanied by cell division, migration, and rearrangement, 
making these two contributions difficult to 
separate~\cite{hohn2015dynamics}; it is thus of general interest
to identify a fixed cellular assembly amenable to precise 
measurements.

\begin{figure}[b!]
\includegraphics[width=1.0\columnwidth]{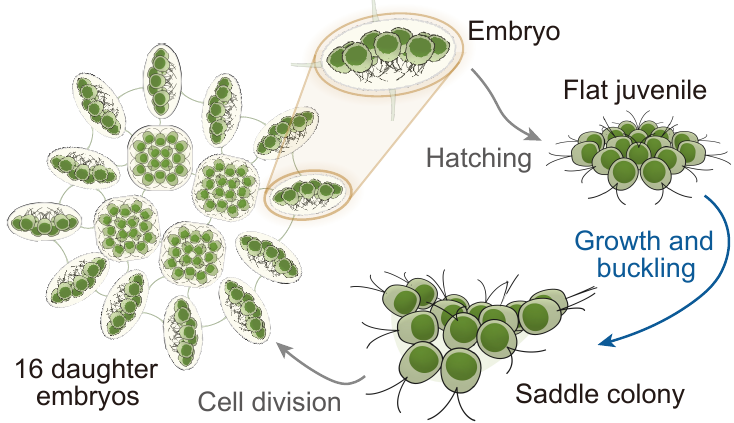}
\caption{\label{fig:1}%
Life cycle of \textit{G. pectorale}. Each
cell divides into a 16-cell daughter embryos which hatch as flat juveniles that grow and 
buckle into a saddle (this work) before dividing again. 
}
\end{figure}

Here we show that the $16$-cell volvocine alga 
\textit{Gonium pectorale}~\cite{hanschen2016gonium,herron2008evolution,de2020motility}, 
whose life cycle is
shown in Fig.~\ref{fig:1}, provides such an assembly.  Once hatched, the juvenile is a flat plate
whose $4$ central and $12$ peripheral cells (``$4+12$") grow while held together with an ECM, eventually buckling.
Later, each cell divides to produce $16$ new $16$-cell progeny, with $12$ in a pinwheel 
arrangement around the central $4$ (Fig. \ref{fig:1}, left).
This cycle provides a minimal setting to examine how 
differential cell growth is transmitted through a mechanically connected ECM: its small, 
fixed architecture permits cell-resolved growth measurements, while micropipette aspiration 
allows continuous, long-term observation of colony shape.

\begin{figure*}[ht!]
\includegraphics[width=\textwidth]{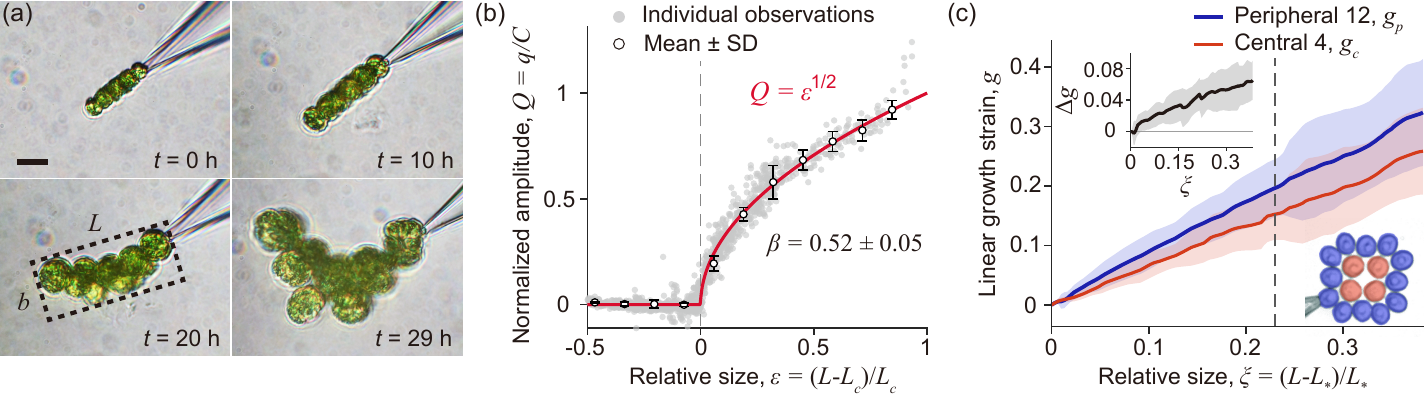}
\caption{\label{fig:2}%
Flat-to-saddle transition of growing \textit{G. pectorale}. (a) Edge-on time-lapse of a 
colony held by a micropipette, with dashed minimum-area bounding rectangle in $t=20$~h 
frame. 
Scale bar: 10~$\mu$m.  (b) Normalized shape amplitude versus relative size for six colonies, where
$C$ is a colony-specific constant; $\beta$ is obtained from per-colony fits of $q\propto \epsilon^\beta$.  
(c) Cumulative linear growth strain of peripheral (blue) and central (red) cells
from $6$ face-on growth records. 
The dashed line marks the mean critical size $\overline L_c$ of
the shape cohort. Upper inset: mismatch
$\Delta g=g_{p}-g_{c}$.
Lower inset: central and peripheral cell groups.}
\end{figure*}

\begin{figure*}[t]
\centering
\includegraphics[width=\textwidth]{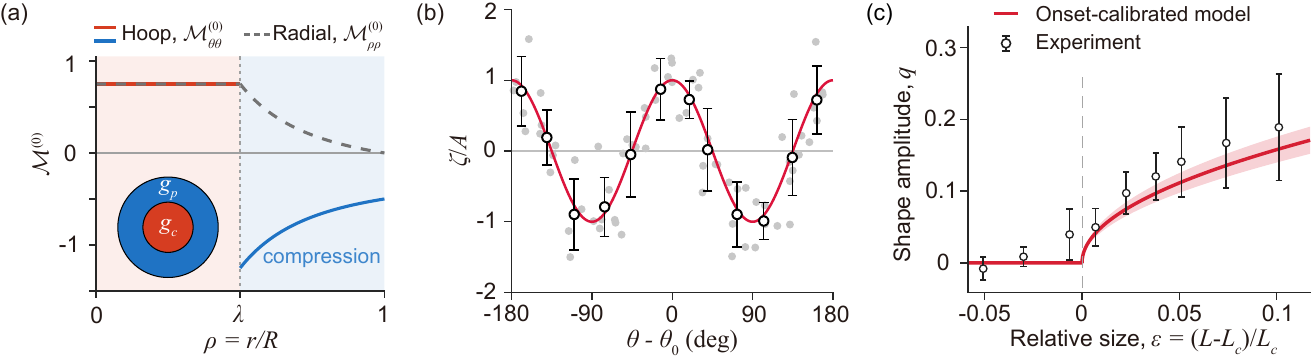}
\caption{\label{fig:3}%
Growth-induced stress, saddle symmetry, and near-onset amplitude. (a) Base-state stress 
resultants ${\mathcal M}^{(0)}$ of the two-region disk for $\lambda=1/2$: radial (dashed) and 
hoop (solid). Inset: growth regions. (b) Peripheral cell heights of $6$ colonies relative 
to their best-fit planes, each rotated by its own $m=2$ phase $\theta_0$ and normalized 
by its fitted amplitude $A$ (grey). Circles show $30^\circ$ bins; red line is the mode 
shape $\cos2(\theta-\theta_0)$. (c) Amplitude predicted from the six measured 
growth curves and independently measured side-view amplitudes, shown over their 
common near-onset range.}
\end{figure*}

Although ``warping" of \textit{Gonium} colonies was noted  
long ago \cite{harper1912structure}, this feature has 
not been quantified 
or explained.  We find that peripheral cells grow faster than 
central ones, and show that a two-region elastic model 
[Fig. \ref{fig:3}(a)] converts this imbalance into hoop compression that selects the saddle mode 
independently, identified by 3D reconstruction. The continuous onset with exponent $0.52\pm0.05$ 
is consistent with the predicted bifurcation. With the measured critical size 
as input, the growth records account for the near-onset amplitude without fitting. The 
growth pattern thus selects the saddle, while the mechanics of the ECM-connected assembly sets how much growth is needed for it to appear: the onset serves as a mechanical assay, rationalizing an effective mechanical thickness associated with the ECM.

\textit{Shape transition}---Wild-type colonies of 
\textit{G. pectorale} (CCAC strain 3275 B) were grown axenically in 
standard \textit{Volvox} medium~\cite{kirk1983protein} at 
$22\,^{\circ}\mathrm{C}$ under $\sim3800~\mathrm{lx}$ illumination on a 
$14~\mathrm{h}/10~\mathrm{h}$ light/dark cycle. Each newly formed $16$-cell 
colony was held via one peripheral cell by gentle 
aspiration onto a polished glass micropipette of $\sim\! 3~\mu\mathrm{m}$ 
inner diameter, mounted on a rotation stage to enable both 
face-on and edge-on 
imaging. 
The transition occurs during early growth, while the ECM appears 
continuous and the cell number and $4+12$ arrangement are fixed (See \S\! IA of Supplemental 
Material \cite{SM} for details).

Figure~\ref{fig:2}(a) shows an initially flat colony developing a large out-of-plane 
deformation over $29~\mathrm{h}$ (Movie~1 in ~\cite{SM}).  
We enclose the colony body in each image by its minimum-area bounding rectangle, 
with short and long sides $b$ and $L$ and aspect ratio $a=b/L$, and let
\begin{equation}
q=a-a_0,
\qquad {\rm and} \qquad
\varepsilon=(L-L_c)/L_c,
\label{eq:observables}
\end{equation}
where $a_0$ is the pre-transition plateau of $a$ and the critical size 
$L_c$ (\S~IB in \cite{SM}) is that at which buckling first
appears. Across $6$
colonies, $\overline L_c=(46.7\pm 8.9)~\mu\mathrm{m}$ 
(where all uncertainties herein are standard deviations).
The amplitude $q$ rises from zero continuously, with no detectable jump at onset. 
Fitting $q\propto\varepsilon^\beta$ to each colony  
gives $\beta=0.52\pm0.05$. A separate square-root fit 
gives a prefactor $C$ for each colony. The normalized amplitude $Q=q/C$ collapses 
the data onto $Q\simeq\sqrt{\varepsilon}$ [Fig.~\ref{fig:2}(b)]. 
The continuous square-root growth of $q$ is consistent with a supercritical 
pitchfork bifurcation, confirmed below by the sign of the cubic term in an 
amplitude equation. 

\textit{Differential growth}---A separate set of $6$ colonies was 
imaged face-on in bright-field $z$ stacks, with each cell's projected area measured 
in its best-focused plane (\S\! IC in \cite{SM}). For each colony, let $S_\alpha$ (for $\alpha=c,p$) denote the mean area 
of the central $4$ or peripheral $12$ cells [Fig.~\ref{fig:2}(c)]. 
We define the regional linear expansion 
and relative size as
\begin{equation}
g_\alpha=\sqrt{S_\alpha/S_\alpha^{\,\mathrm{*}}}-1, \ \ \ {\rm and} \ \ \ 
\xi=(L-L_{\mathrm{*}})/L_{\mathrm{*}},
\label{eq:growth}
\end{equation}
where $ S_\alpha^{\,\mathrm{*}}$ is the reference mean area at $L_{\mathrm{*}}$, 
the smallest colony size covered by all six records. The square root converts the area ratio into a length ratio, as 
the growth tensor in the model describes expansion of line elements.

Figure~\ref{fig:2}(c) shows greater relative expansion of 
peripheral cells and an overall increase in the mismatch $\Delta g=g_{p}-g_{c}$. At the mean critical size $\overline L_c$ obtained from the shape measurements (dashed line), the differential growth is $\Delta g=\mathrm{0.065}\pm\mathrm{0.028}$. Growth thus builds an increasing mismatch across the fixed $4+12$ arrangement, providing a measured input for the elastic analysis.

{\textit{Theory}}---We model the colony as a thin $2$-region 
elastic disk of radius 
$R=L/2$ and thickness $H$, with radial coordinate $r$ and internal division at $r=\lambda R$. Setting 
$\lambda=1/2$ assigns one quarter of the disk area to the center [Fig.~\ref{fig:3}(a)], as in the $4{:}12$ 
partition. Each region grows locally isotropically, by $g_{\mathrm{c}}$ 
or $g_{\mathrm{p}}$. In the mechanical analysis, strains are referenced to each colony's
first recorded planar state $L_0$, with no additional initial mismatch. The sheet has  
uniform 2D Young modulus $K$ ($K=HY$, where $Y$ is the 
3D Young's modulus), 
bending rigidity $D$, Poisson ratio $\nu=1/2$ and free outer boundary.

Differential growth between the central and peripheral regions
leads to longitudinal stresses that result in out-of-plane displacements
$\zeta(r,\theta)$.  Anticipating a perturbative expansion of the
stress tensor $N$ in powers of $\zeta$,
we consider first
the longitudinal stresses $N^{(0)}_{\alpha\beta}$ in a flat disk in the membrane limit 
\cite{landau2012theory}, where the 2D elastic strain tensor $E$ is
related to the displacement ${\bf u}$ and growth rate $g$ via
$E_{\alpha\beta}=(u_{\alpha,\beta}+u_{\beta,\alpha})/2-g\delta_{\alpha\beta}$.
With azimuthal symmetry and radial displacement 
${\bf u}=u(r)\hat{\bf r}$, 
the stress resultants $N^{(0)}_{rr}$ and $N^{(0)}_{\theta\theta}$ 
are given by Eq.~\eqref{Ns} in End Matter,
with $N^{(0)}_{r\theta}=0$.  The equilibrium condition $\partial N^{(0)}_{\alpha\beta}/\partial x_\beta=0$
reduces to $u^{\prime\prime}+u^{\prime}/r-u/r^2=0$, with 
regularity at $r=0$, a free edge at $r=R$,
and continuity in $u$ and $N^{(0)}_{rr}$ at $r=\lambda R$.  
With the rescalings
$N^{(0)}_{\alpha\beta}=K{\cal N}^{(0)}_{\alpha\beta}$ and $\rho=r/R$
we find 
$\mathcal{N}^{(0)}_{\rho\rho}=\mathcal{N}^{(0)}_{\theta\theta}=\Delta g(1-\lambda^2)/2$ in the center and in the periphery ($\lambda<\rho\leq 1$)
\begin{equation}
{\cal N}^{(0)}_{\rho\rho}
=\frac{\Delta g\lambda^2}{2}\left(\rho^{-2}-1\right),\ \  
{\cal N}^{(0)}_{\theta\theta}
=-\frac{\Delta g\lambda^2}{2}(1+\rho^{-2}).
\label{eq:stress}
\end{equation}
As all the nonzero components of ${\cal N}^{(0)}$ are proportional to $\Delta g/2$, we
let ${\cal N}_{\alpha\beta}^{(0)}=(\Delta g/2){\cal M}_{\alpha\beta}^{(0)}$. For $\Delta g >0$ the 
center is under uniform biaxial tension, and although growth is isotropic,
$\mathcal N^{(0)}_{\theta\theta}$ jumps at $\rho=\lambda$ 
and is compressive
throughout the outer annulus [Fig.~\ref{fig:3}(a)]; as the two regions are connected, the
faster-growing periphery cannot expand freely in the circumferential
direction.  

These results derive from an Airy
stress function $\chi$ via $N_{\alpha\beta}=
\epsilon^{\alpha \gamma}\epsilon^{\beta\delta}\partial^2\chi/\partial {x_\gamma}\partial {x_\delta}$,
and scaled $\Phi
=\chi/KR^2$. In the center, we find
$\Phi^{(0)}=\Delta g (1-\lambda^2)
\rho^2/4$, while in the periphery 
$\Phi^{(0)}=(\Delta g\lambda^2/2)(\ln(\rho/\lambda)-\rho^2/2+1/2)$.
Now allowing for non-zero $\zeta$, we set $\Phi=\Phi^{(0)}+\tilde{\Phi}$, 
where $\tilde\Phi$ is the correction. The 
spatially nonuniform growth is accounted for by the flat-state 
stress function $\Phi^{(0)}$. Subtracting this,
the growth-modified F\"oppl--von K\'arm\'an 
equations~\cite{dervaux2008morphogenesis} 
are coupled PDEs for the scaled displacement $w=\zeta/R$ and $\tilde{\Phi}$.  Defining $\Gamma=D/(KR^2)$, they are
\vspace{-3mm}
\begin{subequations}
\begin{align}
\Gamma\nabla^4 w-{\cal N}^{(0)}_{\alpha\beta}\,w_{,\alpha\beta}
&=2[\tilde\Phi,w],\label{eq:fvk1}\\
\nabla^4\tilde\Phi&=-[w,w],\label{eq:fvk2}
\end{align}
\label{eq:fvk}%
\end{subequations}
where $[f,g]=(f_{xx}g_{yy}+f_{yy}g_{xx}-2f_{xy}g_{xy})/2$, and
$[w,w]$ is the Gaussian curvature of the sheet (\S\! IIA in \cite{SM}).

Linearizing Eq.~\eqref{eq:fvk} about the flat state removes the right-hand
side of its first line, which is cubic in $w$ because $\tilde\Phi=O(w^2)$.
The linear problem becomes
\begin{equation}
\nabla^4 w=\eta\mathcal M^{(0)}_{\alpha\beta}\,w_{,\alpha\beta}, 
\qquad \eta=\frac{\Delta g}{2\Gamma}.
\label{eq:control1}
\end{equation}
Eq. \eqref{eq:control1} depends on growth and 
mechanics only through $\eta$; it
must be solved in each region with 
boundary conditions at $\rho=0, \lambda, 1$, and $\eta=\eta_c(m)$ is an 
eigenvalue for modes
$w\propto\phi_m(\rho)\cos(m\theta)$.
For $\lambda=\nu=1/2$, the first few numerically obtained thresholds 
\cite{SM} are $\eta_c(2)=9.16$ (a saddle), 
$\eta_c(3)=19.4$, $\eta_c(4)=33.7$. Axisymmetric  
buckling occurs if the center outgrows the periphery
($\Delta g < 0$). 

In our experiments, we identify the mode from cell-center positions 
reconstructed from the face-on $z$ stacks (\S\! ID and Movie 2 in \cite{SM}). 
The displacements of peripheral cells from a best-fit plane, 
aligned in phase and normalized by amplitude ($\zeta/A$), follow the 
azimuthal variation of the predicted saddle [Fig.~\ref{fig:3}(b)]. The 
significance of this fit is quantified by $P_m$, the fraction 
of the squared 
peripheral height signal explained 
by a separate least-squares fit of mode $m$. Across the $6$ colonies, 
$m=2$ is the largest mode in every colony and $P_2=(73\pm9)\%$. The 
observed spatial symmetry therefore matches the first instability 
generated by the measured growth pattern.

\textit{Onset and amplitude}---Buckling sets in when $\eta=\eta_c(2)$, so the measured critical mismatch $\Delta g_c$ determines 
the onset value of the stiffness ratio via 
$\Gamma_c=\Delta g_c/2\eta_c(2)$. This makes buckling a mechanical assay 
for $\Gamma$. Because growth and shape were recorded in separate 
cohorts, we evaluate each growth curve at the mean critical size 
$\overline L_c$ of the shape cohort. The six resulting estimates 
give $\overline\Gamma_c=(3.5\pm 1.5)\times 10^{-3}$, which we 
discuss below.

Beyond onset, the amplitude is limited by the nonlinearities in Eq.~\eqref{eq:fvk}. Near onset,
$w=\mathcal A\,\phi_c(\rho)\cos 2\theta+\cdots$, where
$\phi_c\equiv\phi_2$ is the critical mode, normalized
so that $\max|\phi_c|=1$, and $\mathcal A$ is the
dimensionless maximum rim deflection. Projecting Eq.~\eqref{eq:fvk} onto this mode gives
\begin{equation}
0=c_1(\eta-\eta_c)\mathcal A-c_3\mathcal A^3+\cdots,
\qquad c_1,c_3>0,
\label{eq:landau}
\end{equation}
which contains only odd powers because \eqref{eq:fvk} is invariant
under $\zeta\to-\zeta$. The cubic coefficient is positive because a saddle
has nonzero Gaussian curvature and therefore cannot form without stretching
the sheet, at an energetic cost of order $\mathcal A^4$ (End Matter). The amplitude 
therefore undergoes a
supercritical pitchfork bifurcation, consistent with the continuous square-root onset
in Fig.~2(b).

Converting $\mathcal A$ to the side-view observable gives 
$q^2\simeq C_g(\Delta g-2\eta_c\Gamma)$ above threshold, where 
$C_g=c_1/(2\Gamma c_3)\simeq1.08$ is calculated from the mode shape and the 
observation geometry (\S\! IIC in \cite{SM}). Over the near-onset window we 
take $\Gamma\simeq\Gamma_c$, so that $2\eta_c\Gamma\simeq\Delta g_c$ and, 
for $\Delta g>\Delta g_c$,
\begin{equation}
q^2\simeq C_g\,(\Delta g-\Delta g_c).
\label{eq:amplitude}
\end{equation}
The growth records fix $\Delta g-\Delta g_c$ and $C_g$ is known, 
so Eq.~\eqref{eq:amplitude} has no adjustable parameters. Since the measured mismatch increases approximately linearly
with size near onset, Eq.~\eqref{eq:amplitude} also predicts
$q\propto\varepsilon^{1/2}$, consistent with the square-root
onset observed in Fig~\ref{fig:2}(b).

We evaluate Eq.~\eqref{eq:amplitude} for each growth record, with 
$\Delta g_c$ taken at $\overline L_c$, and average the amplitudes. 
Figure~\ref{fig:3}(c) compares these predictions with the experimental amplitudes over the common interval, 
$-0.06\lesssim\varepsilon\lesssim 0.12$, with each shape record 
keeping its critical size. The prediction reproduces both the observed 
amplitude scale and its increase with $\varepsilon$ within
the between-colony spread.

With this effective description supported by the amplitude comparison, 
we compare the inferred stiffnesses with a homogeneous-sheet benchmark. 
For an incompressible homogeneous sheet whose thickness equals the cell-body
diameter $H$, $D/K=H^2/9$, which is a purely geometric result. With $\Gamma_c=D/(KR_c^2)$ and 
$R_c=\overline L_c/2$, the colony-averaged $D/K$ is smaller than this 
benchmark by a factor of $\simeq5.4$, implying substantially weaker 
resistance to bending, relative to stretching, than for a homogeneous 
sheet of cell-body thickness. Equivalently, the mechanical 
thickness $h\equiv3(D/K)^{1/2}=3R_c\Gamma_c^{1/2}$, which equals $H$ for
a homogeneous sheet, is $h=(0.42\pm0.10)H$. Thus the buckling onset 
probes a load-bearing thickness less than half the colony thickness 
(the cell diameter $H$).

\textit{Discussion}-- Thin intercellular ECM regions offer a possible explanation for this 
reduced ratio. For an elastic strip of thickness $h$, stretching 
stiffness scales $\sim\!h$ and bending rigidity $\sim\!h^3$, so 
thinning preferentially reduces 
bending resistance. We observe mesh-like ECM 
and elongated tethers after the saddle transition 
(\S III in \cite{SM}). Thus, a nonuniform thickness may already 
be present in the ECM during buckling. A related distinction 
between geometric and effective bending 
thickness is incorporated in microtubule shell models to account for 
deformation in thin interprotofilament 
bridges~\cite{memet2018microtubules}. If bending similarly localizes 
between \textit{Gonium} cells, the effective mechanical thickness would 
depend on these compliant regions, even when cell bodies set the visible 
thickness. Resolving ECM thickness and deformation during the transition 
would directly test this interpretation.

\begin{figure}[t!]
\includegraphics[width=1\columnwidth]{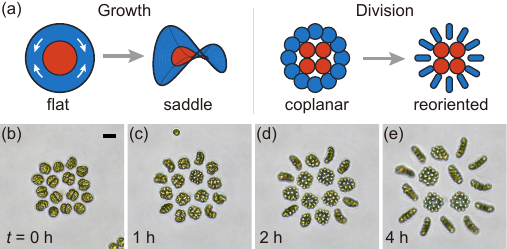}
\caption{\label{fig:4}%
Peripheral crowding relieved out of plane. (a) Schematics of growth and division. (b)--(e) Time-lapse showing peripheral daughter colonies reorienting
out of plane (\S\! IE in \cite{SM}). Scale bar, $10~\mu\mathrm{m}$.
}
\end{figure}

Peripheral crowding offers a possible geometric link between saddle
buckling and the subsequent colony development. Harper 
already noted \cite{harper1912structure} that the $12$ peripheral
cells occupy ``the space which would be more naturally filled by 
ten''. Excess peripheral growth intensifies
this crowding, and the sheet relieves it by bending out of plane. A
similar out-of-plane accommodation reappears at division
(Fig.~\ref{fig:4}): once each cell has formed a 16-cell daughter within
the tethered matrix, the $12$ peripheral daughters stand edge-on in a
ring around the $4$ central ones, which reorient only later. 
Together, these observations point to a common geometric principle: in 
an ECM-tethered colony, peripheral crowding is relieved out-of-plane, 
by collective bending of the sheet during growth and by reorientation 
of the connected daughters at division. This may be an 
adaptive trait that allows a multicellular organism to bear large
multicellular progeny. A quantitative understanding of the late-stage
geometry seen in Fig.~\ref{fig:4} likely lies far
outside the realm of F{\"o}ppl-von Karman theory and would require
a more intrinsic formulation of elastic sheet dynamics of the kind
that has been described recently \cite{cohen2026curvature}.

The comparison with uni- and multicellular relatives highlights a role 
of the ECM. In \textit{Chlamydomonas}, the homologous material is a 
wall around each cell; the daughter cells formed within the 
mother-cell wall are released as independent
individuals~\cite{kubo2009chlamydomonas}, so no shared 
matrix couples their growth. In \textit{Volvox}, embryos develop 
inside glycoprotein vesicles whose controlled enlargement is required 
for inversion; when the vesicle fails to expand, inversion is 
impaired, and releasing the embryo restores
it~\cite{ueki2009controlled}. \textit{Gonium} lies between these 
two: its cells, and later its daughters, are coupled through 
a common ECM, so growth of the parts generates stresses borne by 
the whole. The shape changes reported here are thus a direct 
mechanical consequence of multicellularity, of cells that grow 
while remaining attached. Tuning the growth pattern selects the 
mode, whereas remodelling of the matrix sets how much growth is 
needed for it to appear, offering two distinct routes by which an organism can control its form.

\begin{acknowledgements}
We thank Hester van Wachem for discussions. This work was supported in part by 
the Wellcome Trust Discovery Award 307079/Z/23/Z (ZW \& JYYC), the 
Leverhulme Trust (KL \& SKB)
and the Complex Systems Fund at the University of Cambridge. 
\end{acknowledgements}

\textit{Data Availability-} The data that support the findings of this article are openly available \cite{Zenodo}.

\bibliography{main_refs.bib}

\clearpage

\section{End Matter}
\label{endmatter}

\textit{Base state}---For a flat, axisymmetric colony, the displacement is radial, ${\bf u}=u(r)\hat{\bf r}$, and the elastic strains are
$E_{rr}=u'-g$ and $E_{\theta\theta}=u/r-g$, where $'=d/dr$ and 
$g$ takes the constant value $g_{\mathrm c}$ or
$g_{\mathrm p}$ in the respective regions. The 2D form of Hooke's law 
for the sheet is
\begin{subequations}
\begin{align}
N^{(0)}_{rr}&=\frac{K}{1-\nu^2}\left(E_{rr}+\nu E_{\theta\theta}\right),\\
N^{(0)}_{\theta\theta}&=\frac{K}{1-\nu^2}\left(E_{\theta\theta}+\nu E_{rr}\right).
\end{align}
\label{Ns}%
\end{subequations}
Radial force balance,
$dN^{(0)}_{rr}/dr+(N^{(0)}_{rr}-N^{(0)}_{\theta\theta})/r=0$,
reduces in each region to
$u''+u'/r-u/r^2=0$. Thus,
$u(r)=A_\alpha r+B_\alpha/r$, for $\alpha=\mathrm c,\mathrm p$.

Regularity at the center requires $B_{\mathrm c}=0$, and the
free outer edge requires $N^{(0)}_{rr}(R)=0$, and at the bonded interface
$r=\lambda R$ both $u$ and, by radial force balance, $N^{(0)}_{rr}$
are continuous. These four conditions yield
\begin{equation}
\begin{aligned}
A_{\mathrm c}
&=g_{\mathrm c}
+\frac{1-\nu}{2}(1-\lambda^2)\Delta g,\\
A_{\mathrm p}
&=g_{\mathrm p}
-\frac{1-\nu}{2}\lambda^2\Delta g,\\
B_{\mathrm p}
&=-\frac{1+\nu}{2}\lambda^2R^2\Delta g,
\end{aligned}
\label{eq:em_base_constants}
\end{equation}
where $\Delta g=g_{\mathrm p}-g_{\mathrm c}$. Substituting these
constants into Eqs.~\eqref{Ns} and writing $\rho=r/R$ and dividing by $K$ gives the base stresses $\mathcal N^{(0)}_{\alpha\beta}$ of the main text.

\vspace{0.5cm}
\textit{Linear stability}---For separable solutions to Eq. \eqref{eq:control1}
of the form $w=\phi_m(\rho)\cos(m\theta)$ we define $\mathcal L_m=d^2/d\rho^2+\rho^{-1}d/d\rho-m^2/\rho^2$ and use the result
$\mathcal M^{(0)}_{\rho\theta}=0$ to obtain
\begin{equation}
\mathcal L_m^2\phi_m
=\eta\left[
\mathcal M^{(0)}_{\rho\rho}\phi_m''
+\mathcal M^{(0)}_{\theta\theta}
\left(\frac{\phi_m'}{\rho}-\frac{m^2\phi_m}{\rho^2}\right)
\right].
\label{eq:em_linear_mode}
\end{equation}
Substituting the components of $\mathcal M^{(0)}$ gives
$$ 
\mathcal L_m^2\phi_m=(1-\lambda^2)\,\eta\,\mathcal L_m\phi_m,
$$
in the center, and in the periphery, 
\begin{align}
\mathcal L_m^2\phi_m=\lambda^2\eta\biggl[&(\rho^{-2}-1)\phi_m''\nonumber\\
&-(1+\rho^{-2})\left(\frac{\phi_m'}{\rho}-\frac{m^2\phi_m}{\rho^2}\right)
\biggr].\nonumber
\end{align}
Growth and mechanics thus enter only through $\eta$.

Each regional equation is fourth order, giving four integration constants each 
in the center
and the periphery. These eight conditions are: \textit{(center)} The deflection
$w=\phi_m\cos(m\theta)$ must be smooth and single-valued at the center. For $m\ge2$, this requires
$\phi_m(0)=\phi_m'(0)=0$;
\textit{(interface)} The sheet can neither tear nor fold 
along $\rho=\lambda$: its deflection and slope are continuous. No force or torque is applied along the interface, so the bending moment and transverse force are continuous.
With $D$ and $\nu$ equal in the two regions,
and the radial tension continuous, these four conditions amount to continuity of
$\phi_m$, $\phi_m'$, $\phi_m''$ and $\phi_m'''$ at
$\rho=\lambda$;
\textit{(periphery)} This edge is free, so it carries no bending moment and no transverse force, the latter including the contribution of twisting along the edge. Substituting
$w=\phi_m\cos(m\theta)$ into the standard expressions for these
quantities~\cite{landau2012theory}, at $\rho=1$, gives
\begin{align}
&\phi_m''+\nu\bigl(\phi_m'-m^2\phi_m\bigr)=0,\nonumber \\
&\phi_m'''+\phi_m''-\bigl[1+(2-\nu)m^2\bigr]\phi_m'
+(3-\nu)m^2\phi_m=0.\nonumber
\end{align}

For each $m\ge2$, the eight conditions allow a nonzero deflection only at particular values of $\eta$. The smallest positive value
defines $\eta_c(m)$. For $\lambda=\nu=1/2$,
numerical solution (Chebyshev collocation) gives $\eta_c(2)=9.16$, $\eta_c(3)=19.4$ and $\eta_c(4)=33.7$, with the thresholds increasing monotonically up to
$m=10$. After removing rigid translation and tilt, no positive buckling
eigenvalue was found for $m=0$ or $1$ (\S\! IIB in \cite{SM}). The $m=2$ saddle is therefore the first unstable mode.

\vspace{0.5cm}
\textit{Amplitude equation}---
Near onset we expand in the amplitude $\mathcal A$ of the critical mode $w_1=\phi_c(\rho)\cos2\theta$,
normalized by $\phi_c(1)=1$. In an expansion in a formal small parameter $\epsilon$, the leading deflection would be $\epsilon A w_1$, so that $\mathcal A=\epsilon A$. Equations~\eqref{eq:fvk} are invariant under
$w\to-w$ with $\tilde\Phi$ unchanged, so the deflection contains only odd
powers of $\mathcal A$ and the induced stress function only even powers:
\begin{subequations}
\label{eq:em_amplitude_expansion}
\begin{align}
w&=\mathcal A w_1+\mathcal A^3w_3+O(\mathcal A^5),\\
\tilde\Phi&=\mathcal A^2\tilde\Phi_2+O(\mathcal A^4).
\end{align}
\end{subequations}
At fixed $\Gamma$ the
critical mismatch is $\Delta g_c=2\eta_c\Gamma$. Since the base stress
is proportional to $\Delta g$,
we have $\mathcal N^{(0)}_{\alpha\beta}=(\eta/\eta_c)
\mathcal N^{(0)}_{\alpha\beta,c}$, where $\mathcal N^{(0)}_{\alpha\beta,c}$
is its value at onset. We take $(\eta-\eta_c)/\eta_c=\mu\mathcal A^2+O(\mathcal A^4)$ and determine $\mu$ at order $\mathcal A^3$.

At $\mathcal O({\mathcal A})$, Eq.~\eqref{eq:fvk1} gives the linear
problem $\Gamma\nabla^4w_1=\mathcal N^{(0)}_{\alpha\beta,c}w_{1,\alpha\beta}$, which 
determines the mode shape, leaving $\mathcal A$ unknown.
At $\mathcal O({\mathcal A}^2)$, Eq.~\eqref{eq:fvk2} gives
\begin{equation}
\nabla^4\tilde\Phi_2=-[w_1,w_1],
\label{eq:em_quadratic_stress}
\end{equation}
so the Gaussian curvature generates an induced stress function of
$\mathcal O({\mathcal A}^2)$. The function $\tilde\Phi_2$ is smooth at the
center and continuous with its first three radial derivatives at
$\rho=\lambda$. At the traction-free outer edge $\rho=1$, we choose the stress function to satisfy $\tilde\Phi_2=\partial_\rho\tilde\Phi_2=0$. At $\mathcal O({\mathcal A}^3)$, Eq.~\eqref{eq:fvk1} divided by $\Gamma$ gives
\begin{equation}
\nabla^4w_3-\eta_c\mathcal M^{(0)}_{\alpha\beta}w_{3,\alpha\beta}=\mu\eta_c\,\mathcal M^{(0)}_{\alpha\beta}w_{1,\alpha\beta}+2[\tilde\Phi_2,w_1]/\Gamma.\nonumber
\label{eq:em_cubic_order}
\end{equation}
The operator on the left hand side is
self-adjoint under the boundary and matching conditions and annihilates
$w_1$. Multiplying by $w_1$ and integrating over the unit
disk, with $dS=\rho\,d\rho\,d\theta$, eliminates $w_3$: the
right side must have zero projection onto $w_1$. Integrating the 
prestress term by parts and using
$\eta-\eta_c\simeq\eta_c\mu\mathcal A^2$, we find
\begin{equation}
\begin{aligned}
0={}&-\Gamma(\eta-\eta_c)\mathcal A
\int \mathcal M^{(0)}_{\alpha\beta}
w_{1,\alpha}w_{1,\beta}\,dS\\
&+2\mathcal A^3\int w_1[\tilde\Phi_2,w_1]\,dS.
\end{aligned}
\label{eq:em_projected_balance}
\end{equation}

Comparing with Eq.~\eqref{eq:landau}, we obtain
\begin{equation}
\begin{aligned}
c_1&=-\Gamma
\int\mathcal M^{(0)}_{\alpha\beta}
w_{1,\alpha}w_{1,\beta}\,dS,\\
c_3&=-2\int w_1[\tilde\Phi_2,w_1]\,dS.
\end{aligned}
\label{eq:em_amplitude_coefficients}
\end{equation}

For $\lambda=\nu=1/2$, $c_1/2\Gamma\simeq0.680$ and $c_3\simeq0.630$; since
$q\simeq|\mathcal A|$ in the edge-on geometry~\cite{SM}, this gives
$C_g=c_1/(2\Gamma c_3)\simeq1.08$.

With the in-plane
displacement relaxed, the elastic energy relative to the flat state is,
in units of $KR^2$,
\begin{equation}
\mathcal E-\mathcal E_0=-\frac{c_1}{2}(\eta-\eta_c)\mathcal A^2
+\frac{c_3}{4}\mathcal A^4+O(\mathcal A^6),
\label{eq:em_amplitude_energy}
\end{equation}
whose stationarity condition is Eq.~\eqref{eq:landau}. The quadratic
term combines the bending cost with the energy released from the
prestress, which balance at onset; the quartic term is the stretching
cost of the saddle. 
The flat state
is stable below onset and unstable above it, where two minima
$\mathcal A=\pm[(c_1/c_3)(\eta-\eta_c)]^{1/2}$, corresponding to
opposite deflections, emerge continuously: a supercritical pitchfork 
bifurcation.


\clearpage
\newpage
\setcounter{equation}{0}
\setcounter{figure}{0}
\setcounter{table}{0}
\setcounter{page}{1}
\makeatletter

\renewcommand{\theequation}{S\arabic{equation}}
\renewcommand{\thefigure}{S\arabic{figure}}
\renewcommand{\bibnumfmt}[1]{[S#1]}
\renewcommand{\citenumfont}[1]{S#1}

\renewcommand{\theHequation}{SM.\arabic{equation}}
\renewcommand{\theHfigure}{SM.\arabic{figure}}
\renewcommand{\theHtable}{SM.\arabic{table}}

\widetext
\begin{bibunit}[apsrev4-2]

\section{Supplemental Material}
\setcounter{equation}{0}
\setcounter{figure}{0}
\setcounter{table}{0}
\renewcommand{\theequation}{S\arabic{equation}}
\renewcommand{\thefigure}{S\arabic{figure}}
\renewcommand{\thetable}{S\arabic{table}}

\setcounter{secnumdepth}{3}
\renewcommand{\thesection}{\Roman{section}}
\renewcommand{\thesubsection}{\Alph{subsection}}
\renewcommand{\thesubsubsection}{\arabic{subsubsection}}
\section{EXPERIMENTAL METHODS}

\subsection{Culture, micropipette holding, and imaging}
\label{sec:sm_culture_imaging}

Wild-type colonies of \textit{Gonium pectorale} (strain CCAC 3275 B, Central Collection of Algal Cultures) were grown axenically in standard \textit{Volvox} medium (SVM)~\cite{kirk1983protein}. Cultures were maintained at $22\,^{\circ}\mathrm{C}$ under approximately $3800~\mathrm{lx}$ illumination on a $14$-h/$10$-h light/dark cycle. Newly released sixteen-cell colonies were selected for recordings, which followed growth toward the next round of cell division. The transition analyzed in the main text occurred while the extracellular matrix (ECM) appeared continuous between cells and the cell number and $4+12$ arrangement remained unchanged. Later developmental changes in the ECM are described separately in \S\!~\ref{sec:ecm}.

Micropipettes were pulled from borosilicate glass capillaries (Sutter Instruments; outer diameter $1.0~\mathrm{mm}$, inner diameter $0.58~\mathrm{mm}$) using a P-97 puller (Sutter Instruments). The tips were polished using a DMF1000 microforge controller (World Precision Instruments) to an inner diameter of approximately $3~\mu\mathrm{m}$. Colonies were transferred to a Petri dish and held individually by gentle aspiration, with suction supplied manually by a $10$-mL syringe. The micropipettes were mounted on micromanipulators (Patchstar, Scientifica, UK) carried by a custom rotation stage, allowing face-on or edge-on orientation relative to the optical axis. A layer of mineral oil covered the medium to limit evaporation and the associated convective flows.

The pipette aperture was smaller than a single cell, restricting direct contact to only one peripheral cell. The contacted cell continued to grow and subsequently divided, as did its neighbors. Saddle-shaped colonies were also observed during free swimming. In the continuum model, the localized pipette contact is neglected and the outer boundary is taken to be mechanically free.

Edge-on time-lapse recordings of six colonies were acquired at $1$-min intervals on a Zeiss Axiovert 200M inverted microscope with a $40\times$ LD Achroplan objective and a Nikon Z6 II camera. These recordings resolved the evolving colony-body outline used to quantify the flat-to-saddle transition.

A separate set of six colonies was imaged face-on on a Zeiss Axio Observer Z1 inverted microscope with a $40\times$ LD Plan-Neofluar objective and an Axiocam 712 color camera. Bright-field focal series were acquired at $ \sim5$-min intervals over an axial range of approximately $200~\mu\mathrm{m}$ with a spacing of $2~\mu\mathrm{m}$ between planes. Each cell was measured in the plane where its boundary was most sharply resolved. These series provided both the projected cell areas used to measure regional growth and the axial positions used in the three-dimensional reconstruction. The corresponding analysis procedures are described in the following subsections.

\subsection{Shape analysis and critical size}
\label{sec:shape}

The colony-body outline was segmented from each edge-on image by intensity thresholding, excluding the micropipette and flagella. Its minimum-area bounding rectangle has long and short sides $L$ and $b$, giving the aspect ratio $a=b/L$. The pretransition plateau $a_0$ was
estimated from the initial planar portion of each record, and $q=a-a_0$
and $\varepsilon=(L-L_c)/L_c$ are defined as in Eq.~(1) of the main text.

\begin{figure}[h]
\centering
\includegraphics[width=0.7\linewidth]{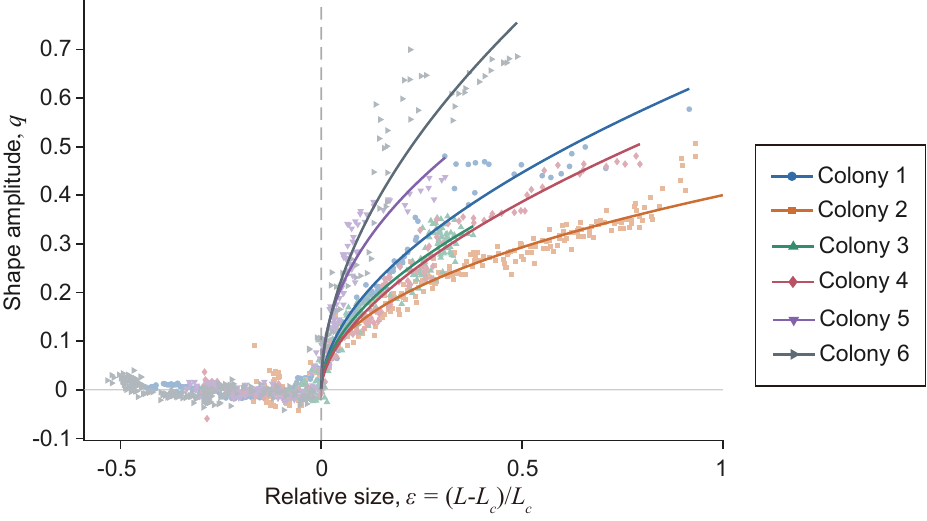}
\caption{Individual shape records. Shape amplitude $q$ versus relative
size $\varepsilon$ for the six edge-on colonies. Lines show the power-law fits for $\varepsilon>0$.}
\label{fig:shape_individual}
\end{figure}

For each colony, $L_c$ and $\beta$ were individually obtained by a least-squares fit of
$q\propto\varepsilon^\beta$ for $\varepsilon>0$, with $q=0$ below onset,
to the record before cell division. Across the six colonies,
$\overline L_c=(46.7\pm8.9)~\mu$m and $\beta=0.52\pm0.05$ (mean $\pm$ SD).
Figure~\ref{fig:shape_individual} shows the six records with their fits:
each colony's own $L_c$ places its onset at $\varepsilon=0$, while the
amplitudes differ between colonies. For the collapse in Fig.~2(b) of the
main text, a separate fit of $q=C\sqrt{\varepsilon}$ with $L_c$ fixed
gives a prefactor $C$ for each colony, and $Q=q/C$. The amplitude
comparison in main text Fig.~3(c) uses $q$ itself.

\subsection{Cell growth and regional mismatch}
\label{sec:growth}

The projected cell areas were measured from the face-on focal series described in \S\!~\ref{sec:sm_culture_imaging}, each cell in its best-focused plane,
and cells were assigned to the central four ($c$) or peripheral
twelve ($p$) according to their positions in the fixed arrangement. For each colony, $S_\alpha(L)$ is the mean cell area of
region $\alpha$ ($\alpha=c,p$) at colony size $L$. The regional linear expansion and relative size $g_\alpha$ and $\xi$ are defined
by Eq.~(2) of the main text, with $\Delta g=g_p-g_c$.
The reference mean areas $S_\alpha^*$ at $L_*$, the smallest colony size
covered by all six records, are obtained by linear interpolation of each record.

\begin{figure}[h]
\centering
\includegraphics[width=0.8\linewidth]{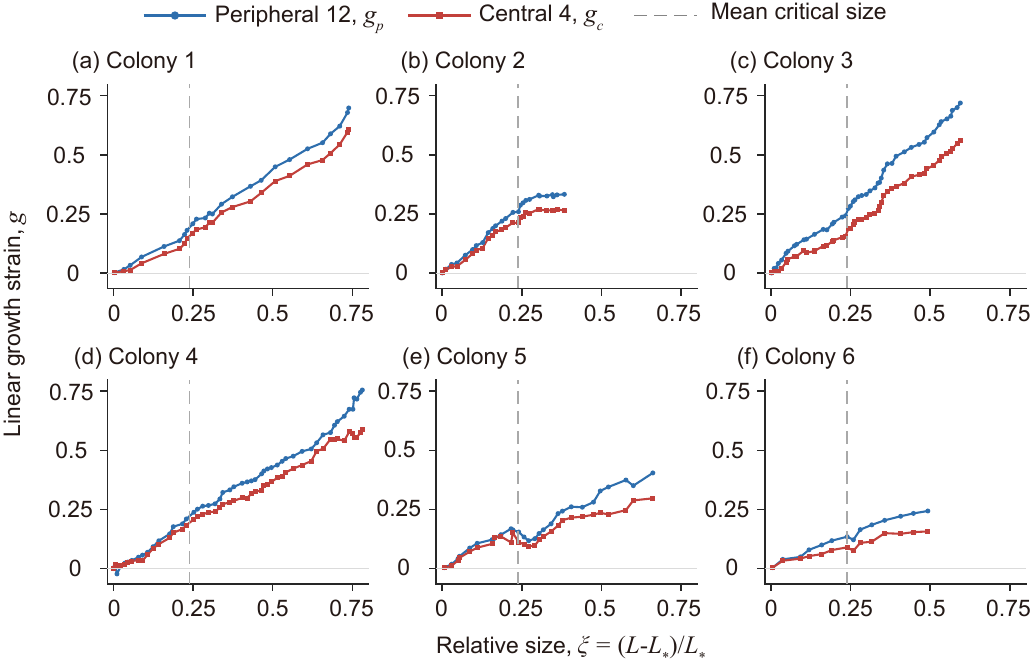}
\caption{Regional growth in individual colonies. (a)--(f) Cumulative linear expansion of the twelve peripheral ($g_p$, blue) and four central ($g_c$, red) cells in each of the six face-on records, versus $\xi=(L-L_*)/L_*$. Dashed lines mark $\overline L_c$. The averages in Fig.~2(c) of the main text use only the range shared by all six
colonies.}
\label{fig:growthSI}
\end{figure}

Figure~\ref{fig:growthSI} shows the individual growth curves. For the
averages in Fig.~2(c) of the main text, each colony's curves are
interpolated onto a common grid in $\xi$, over the range covered by all
six records, and averaged with equal weight; $\Delta g$ is computed
within each colony before averaging, and the bands show the SD across
colonies.

\subsection{Three-dimensional reconstruction}
\label{sec:recons}

Cell-center positions were reconstructed from the face-on focal series
of \S\!~\ref{sec:sm_culture_imaging}. For each cell, its center in the
best-focused image gave the lateral coordinates, and the corresponding
focal plane gave the axial coordinate. These coordinates were used as centroids
of the sixteen cells in a least squares fit to generate a plane for the entire colony.
We denote the heights relative to
this plane by $\zeta_j$ and the azimuths about the colony centroid by
$\theta_j$.

Each colony was analyzed separately using its twelve peripheral cells.
For each $m=1,\ldots,6$, a separate least-squares fit at the measured
azimuths gave
\begin{equation}
\widehat\zeta_m(\theta)=u_m\cos(m\theta)+v_m\sin(m\theta),
\label{eq:recons_harmonic}
\end{equation}
where the hat denotes the fitted height. For $m=0$, the fitted value was
the mean peripheral height. The single-mode explained fraction was
defined as
\begin{equation}
P_m=\frac{\displaystyle\sum_{j\in\mathrm{p}}
\widehat\zeta_m(\theta_j)^2}
{\displaystyle\sum_{j\in\mathrm{p}}\zeta_j^2}.
\label{eq:recons_explained}
\end{equation}
Plane subtraction removes an overall height offset and rigid tilt from
all sixteen cells; it does not force $P_0$ or $P_1$ of the peripheral
subset to vanish. Because the measured azimuths are not uniformly
spaced, the separately fitted harmonics are generally not orthogonal,
and their explained fractions need not sum to unity.

\begin{figure}[hb!]
\centering
\includegraphics[width=0.85\linewidth]{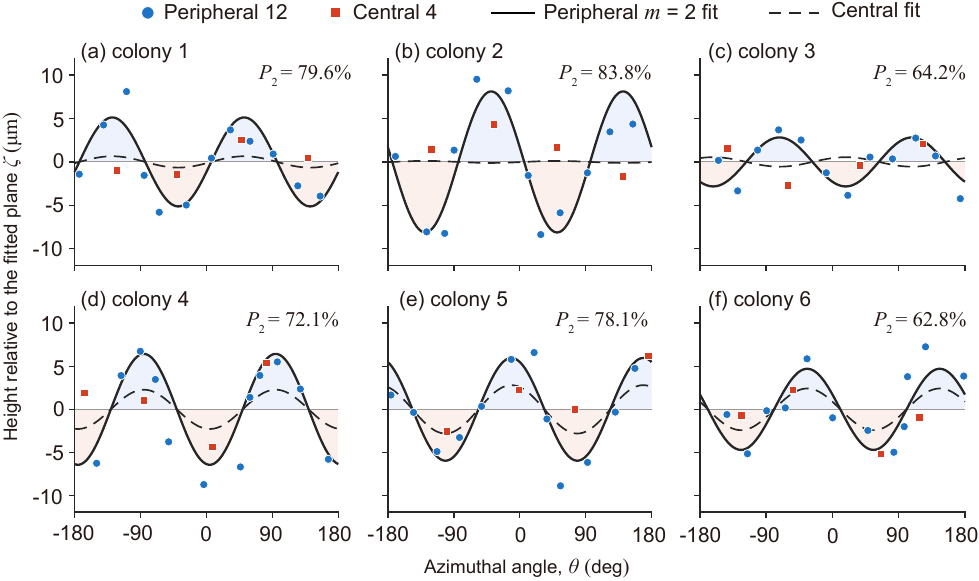}
\caption{Saddle profiles of six individual colonies. Cell-center heights relative to a plane fitted to all sixteen cells are plotted against azimuth. Blue circles: twelve peripheral cells; red squares: four central cells. Solid curves show the peripheral $m=2$ fits; dashed curves show central fits with their phase fixed by the peripheral fit.}
\label{fig:mode_individual}
\end{figure}

For each colony, the peripheral $m=2$ fit was written as
$A\cos[2(\theta-\theta_0)]$, with $A=(u_2^2+v_2^2)^{1/2}$.
Figure~\ref{fig:mode_individual} shows the unnormalized heights of all sixteen cells and the individual fits. The central heights were fitted with the same $m=2$ phase as the periphery, allowing only the amplitude to vary. Central
cells therefore enter the plane fit but are excluded from the
peripheral harmonic fits.

For Fig.~3(b) of the main text, each colony's peripheral data set was
plotted as $\zeta_j/A$ against $\theta_j-\theta_0$.
These normalized peripheral data sets were pooled across all colonies
and displayed as individual points and in $30^\circ$ bins. Central
cells were not included in this figure.

\begin{figure}[h]
\centering
\includegraphics[width=0.45\linewidth]{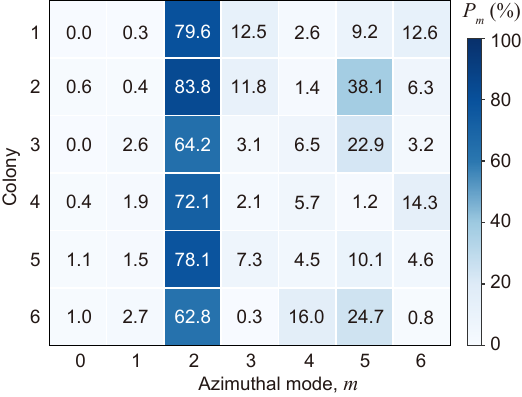}
\caption{Azimuthal mode content across six colonies. Rows correspond to colonies and columns to separately fitted modes $m=0,\ldots,6$. Color and numbers show $P_m$ as a percentage of the squared peripheral height signal explained by each mode after subtraction of the colony-wide best-fit plane.}
\label{fig:mode_summary}
\end{figure}

Figure~\ref{fig:mode_summary} compares $P_m$ across the six colonies.
The $m=2$ fit accounts for the largest fraction in every colony, with
$P_2=(73\pm9)\%$ (mean $\pm$ SD across colonies). This establishes the
reproducibility of the saddle symmetry using cell-center positions
independently of the outline-based amplitude $q$.

\subsection{Division imaging on the chamber floor}
\label{sec:division imaging}

For the division sequence in Fig.~4 of the main text, colonies were immobilized on the floor of a sample chamber rather than held by a micropipette, which is not suited to the late stages of the cycle. Once the 16 daughter colonies have formed [Fig.~1 of the main text], they are linked to one another only through compliant ECM tethers
(\S\!~\ref{sec:ecm}), which cannot maintain their spatial arrangement;
held by a micropipette, the daughters drift out of the focal plane and the field of view. Colonies were instead immobilized by light-induced flagellar adhesion, as reported for
\textit{Chlamydomonas}~\cite{kreis2018adhesion}. The chamber was built
from a glass slide and a coverslip separated by double-sided tape,
giving a depth of $\sim 2$~mm. Under illumination, the flagella of
colonies settling on the glass adhere to it, holding each colony in place for long-term recording. Adhesion keeps the colony flat against the substrate, so these recordings are used only to follow the arrangement of the daughter colonies, not the growth-driven saddle described in the main text. Colonies remained within the field of view during division, allowing daughter formation and reorientation to be
followed continuously. Bright-field time-lapse recordings were acquired
at $1$-min intervals with the microscope, objective, and camera described
in \S\!~\ref{sec:sm_culture_imaging}.

\section{ELASTIC MODEL}
\label{sec:continuum}

\subsection{Incremental equations}
\label{sec:sm_fvk}

We use the geometry, growth field and effective coefficients of the main text. Growth is slow compared with elastic relaxation, so the sheet is in mechanical equilibrium for the current $g$; growth is uniform across the
thickness and imposes no preferred curvature. With
lengths in units of $R$, $w=\zeta/R$, $\Phi=\chi/(KR^2)$ and
$\Gamma=D/(KR^2)$ as in the main text, the growth-modified F\"oppl--von K\'arm\'an
equations~\cite{dervaux2008morphogenesis} for the total stress function are
\begin{subequations}
\label{eq:sm_fvk_total}
\begin{align}
\Gamma\nabla^4 w&=2[\Phi,w],\\
\nabla^4\Phi&=\psi-[w,w],
\end{align}
\end{subequations}
where $[f,g]=(f_{,xx}g_{,yy}+f_{,yy}g_{,xx}-2f_{,xy}g_{,xy})/2$, so that $[w,w]$ is the (scaled) Gaussian curvature of the sheet, and $\psi$
is the Gaussian curvature prescribed by growth. Because uniform growth
merely rescales each region, $\psi$ vanishes within the center and the
periphery and is localized at the interface $\rho=\lambda$.

The flat state $w=0$ satisfies the first of Eqs.~\eqref{eq:sm_fvk_total}
identically, while the second reduces to $\nabla^4\Phi^{(0)}=\psi$. Rather than solving this equation, the End Matter obtains the
corresponding stresses $\mathcal N^{(0)}_{\alpha\beta}$ directly from
the displacement. The Airy representation, $\mathcal N_{xx}=\Phi_{,yy}$, $\mathcal N_{yy}=\Phi_{,xx}$
and $\mathcal N_{xy}=-\Phi_{,xy}$, gives
\begin{equation}
2[\Phi^{(0)},w]=\mathcal N^{(0)}_{\alpha\beta}\,w_{,\alpha\beta}.
\label{eq:sm_base_bracket}
\end{equation}

We compare the buckled sheet with a flat reference state at the
same prescribed growth. Writing $\Phi=\Phi^{(0)}+\tilde\Phi$
and subtracting $\nabla^4\Phi^{(0)}=\psi$ from the second of
Eqs.~\eqref{eq:sm_fvk_total} cancels $\psi$.
Using Eq.~\eqref{eq:sm_base_bracket} in the first equation then gives
the incremental equations of the main text:
\begin{subequations}
\label{eq:sm_fvk}
\begin{align}
\Gamma\nabla^4 w
-\mathcal N^{(0)}_{\alpha\beta}\,w_{,\alpha\beta}
&=2[\tilde\Phi,w],\\
\nabla^4\tilde\Phi&=-[w,w].
\end{align}
\end{subequations}
The growth source $\psi$ cancels from the incremental compatibility
equation, while differential growth remains through the prestress
$\mathcal N^{(0)}_{\alpha\beta}$.

\subsection{Mode selection}
\label{sec:mode_selection}

We solved the radial mode equation with the interface and free-edge
conditions of the End Matter for $m=0,\ldots,10$ at $\lambda=\nu=1/2$.
At the center, smoothness requires $\phi_0'(0)=\phi_0'''(0)=0$ for $m=0$
and $\phi_1(0)=\phi_1''(0)=0$ for $m=1$; the conditions for $m\ge2$ are given in the End Matter.

After excluding rigid translation ($m=0$, constant height) and
rigid tilt ($m=1$, $w\propto\rho\cos\theta$, a tilted plane), no positive buckling eigenvalue was found for $m=0$ or $1$. Over $2\le m\le10$, the thresholds in Table~\ref{tab:sm_mode_thresholds} increase monotonically with $m$.
This trend is consistent with the leading $m^4$ angular bending
term, compared with the $m^2$ hoop-compression term. The lowest threshold, $\eta_c(2)=9.16$, corresponds to the observed saddle. Increasing the numerical resolution did not change the
reported digits.

\begin{table}[h]
\caption{\label{tab:sm_mode_thresholds}
Smallest positive eigenvalue $\eta_c(m)$ for $\lambda=\nu=1/2$. Dashes indicate that none was found after removing rigid translation and tilt.}
\begin{ruledtabular}
\begin{tabular}{lccccccccccc}
$m$ & 0 & 1 & 2 & 3 & 4 & 5 & 6 & 7 & 8 & 9 & 10\\
\hline
$\eta_c(m)$ & --- & --- & 9.16 & 19.4 & 33.7 & 52.1 & 74.6 & 101 & 132 & 166 & 205\\
\end{tabular}
\end{ruledtabular}
\end{table}

\subsection{Amplitude prediction and comparison}

The critical profile reaches its largest magnitude at the rim, where
$\phi_c(1)=1$. The maximum and minimum midsurface heights are thus
$R|\mathcal A|$ and $-R|\mathcal A|$ to leading order, giving a total
height range $2R|\mathcal A|$. Projection along any direction in the
original plane of the colony retains this range, irrespective of the
azimuthal viewing direction.

For the outline measurement, we retain the measured planar baseline
$a_0$ and take the in-plane span to be $L=2R$ to leading order. The
short side of the enclosing rectangle is then
\begin{equation}
b\simeq a_0L+2R|\mathcal A|,
\qquad q=\frac{b}{L}-a_0\simeq|\mathcal A|,
\label{eq:sm_side_view_conversion}
\end{equation}
and the amplitude equation gives, above onset,
\begin{equation}
q^2\simeq C_g(\Delta g-2\eta_c\Gamma),
\qquad C_g=\frac{c_1}{2\Gamma c_3}\simeq1.08.
\label{eq:sm_observable_amplitude}
\end{equation}

The amplitude comparison assumes $\Gamma\simeq\Gamma_c$ over the
near-onset window. Growth and shape were recorded in separate cohorts,
so the mean critical size $\overline L_c$ from the side-view measurements
is used as the onset input for all six face-on growth records. For each
fitted mismatch curve $\Delta g_i(L)$, we evaluate its critical mismatch
$\Delta g_c=\Delta g_i(\overline L_c)=2\eta_c\Gamma_c$ and calculate the
amplitude from Eq.~\eqref{eq:sm_observable_amplitude}. The six predicted amplitudes are then averaged to give the prediction shown in Fig.~3(c) of the main text. The geometric conversion introduces no additional fitted prefactor.

\section{ECM REMODELLING IN LATER STAGES}
\label{sec:ecm}

\begin{figure}[h]
\centering
\includegraphics[width=0.55\linewidth]{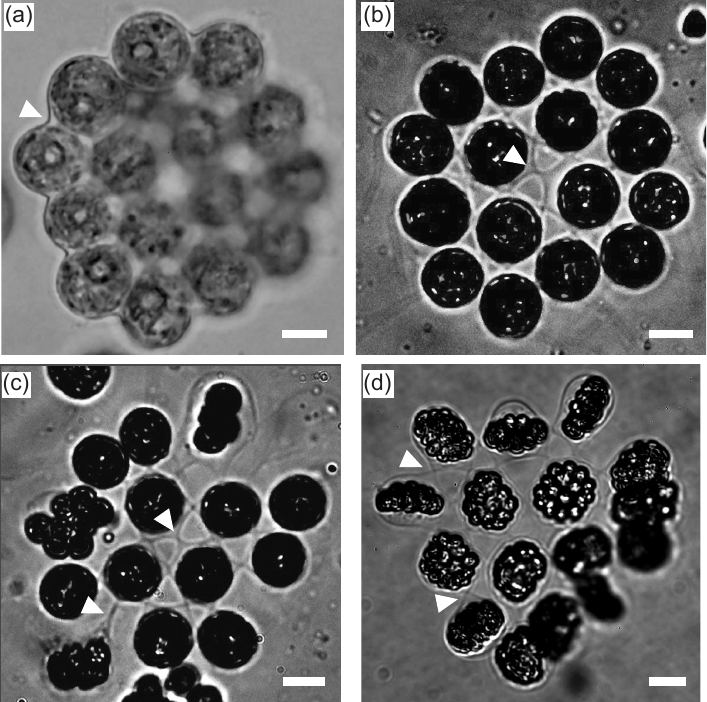}
\caption{Representative ECM morphologies during \textit{Gonium pectorale} development. (a) Colony imaged after Alcian Blue staining; the arrowhead marks the common outer envelope. (b) Mesh-like intercellular matrix, visible between the central cells (arrowhead). (c) Colony containing dividing cells, showing a mesh-like matrix (upper arrowhead) and an elongated connection (lower arrowhead). (d) Developing daughter colonies linked by tether-like structures (arrowheads). The panels show different colonies and do not form a time-lapse sequence. Brightness and contrast were adjusted to improve matrix visibility; images are displayed in grayscale. Scale bars: $10~\mu\mathrm{m}$ in (a)--(c) and $20~\mu\mathrm{m}$ in (d).}
\label{fig:ecm}
\end{figure}

Figure~\ref{fig:ecm} shows ECM morphology at different stages of the developmental cycle summarized in Fig.~1 of the main text. The panels show different colonies rather than successive frames of one colony. In an early colony, Alcian Blue staining reveals a common envelope surrounding the cells [Fig.~\ref{fig:ecm}(a)]. Later in growth, a mesh-like intercellular matrix is visible between the cell bodies, including the four central cells [Fig.~\ref{fig:ecm}(b)]. A colony containing cells that have begun division shows both this mesh-like pattern and elongated intercellular connections [Fig.~\ref{fig:ecm}(c)]. At a more advanced stage, developing daughter colonies remain linked by slender, tether-like structures [Fig.~\ref{fig:ecm}(d)]. During subsequent development, peripheral
daughter colonies turn out of plane, giving the arrangement shown
in Fig.~4 of the main text.

These observations motivate a structural interpretation of the reduced effective bending-to-stretching ratio $\Gamma$ inferred in the main text. The saddle transition analyzed quantitatively occurs during the earlier growth stage, when the matrix appears continuous. At this stage, however, the matrix may already vary in thickness, with thinner regions between neighboring cells. These thin connections could bend more easily while still transmitting in-plane forces, reducing the colony's resistance
to bending relative to stretching. The later mesh and tethers provide morphological support for this possibility. 

\section{SUPPLEMENTAL MOVIE CAPTIONS}

\textbf{Supplementary Movie 1}: Flat-to-saddle transition viewed from the side. Edge-on time-lapse recording of a sixteen-cell \textit{Gonium pectorale} colony held by gentle micropipette aspiration. The colony grows continuously while retaining its 4+12 cell arrangement and, without further cell division, develops from an initially nearly planar configuration into a pronounced out-of-plane shape over approximately 29 h.

\textbf{Supplementary Movie 2}: Axial bright-field scan through a saddle-shaped colony. The focal plane is stepped through the colony at 2-$\mu$m intervals, so that cells at different axial positions come successively into focus, illustrating the measurements underlying the three-dimensional reconstruction.

\textbf{Supplementary Movie 3}: Division of a colony on the chamber floor, corresponding to Fig.~4(b)--(e) of the main text. The colony is immobilized by light-induced flagellar adhesion and remains flat
(\S\!~\ref{sec:division imaging}). Each of the sixteen cells forms a sixteen-cell daughter colony; the twelve peripheral daughters come to
stand edge-on in a ring around the four central ones. Bright-field time-lapse recorded at 1-min intervals. Scale bar, $10~\mu\mathrm{m}$.

\putbib[SI_refs]
\end{bibunit}
\end{document}